# One-Step Self-Assembling Lipopolymer Coatings Enable Direct Longitudinal Biosensing in Native Biological Matrices

*Pascal Rüedi[1], Sarina Seiter[1], Stephan Handschin[2], Chiara Lombardo[1], Jaime Ortega Arroyo[1,*] and Romain Quidant[1]*

[1]Nanophotonics Systems Laboratory, Department of Mechanical and Process Engineering, ETH Zurich, Zurich, Switzerland

[2]Scientific Center for Optical and Electron Microscopy (ScopeM), ETH Zurich, Zurich, Switzerland

*Corresponding author: jarroyo@ethz.ch

Funding: Swiss National Science Foundation grant 207485

Keywords: surface chemistry, label-free imaging, extracellular vesicles, biosensing, supported lipid bilayers, lipid-based nanocarriers

**Abstract**

Sample-driven cross-reactivity remains a fundamental bottleneck for sensitive and selective biosensing in complex biological environments, particularly in microfluidic systems where high surface-to-volumes ratios amplify non-specific interactions. Here, we address this challenge with a one-step, biomimetic surface functionalization strategy based on engineered lipid-based nanocarriers that self-assemble within minutes into immunoassay-ready, antifouling interfaces that couple pegylated-lipid passivation with biorthogonal, site-specific probe conjugation. This approach enables rapid, tuneable, and reproducible surface functionalization, which suppresses sample-specific cross-reactivity even in undiluted serum, and supports direct, label-free digital biosensing assays directly in native biological environments. When integrated into an optofluidic platform combining digital in-line holography and microfluidics, this approach allows time-resolved quantification of tetraspanin-positive extracellular vesicle secretion dynamics from two-dimensional and three-dimensional lung cancer cell models using 50 µL of sample within 10 min, without isolation or purification. Because the presented surface functionalization proceeds via self-assembly rather than covalent surface modification, this one-step modular-approach is intrinsically decoupled from device fabrication and applies to both glass and PDMS, coating the entire microfluidic channel, thereby establishing a general route for integrating biomimetic nanostructures into robust, high-performance immunoassays, advancing biosensing beyond end-point measurements, and towards rapid and continuous real-time monitoring.

## 1. Introduction

One of the most important features of a biosensor is its selectivity to the target molecule, with reagent- and sample-driven cross-reactivity often representing the bottlenecks in sensitivity and applicability across point-of-care and laboratory settings.[1] Whilst reagent-driven cross-reactivity primarily affects multiplexed assays due to the non-specific interactions between analyte, capture and detection probes,[2,3] sample-driven cross-reactivity is universal to all biosensing platforms as it involves the inherent non-specific interactions between non-target analytes with the sensor surfaces. These sample-driven nonspecific interactions compromise assay reproducibility, mask the target-specific signals, lower the sensor's dynamic range and raise the limit of detection (LOD). [4] Microfluidic technology, though a key enabler of high-performance biosensors through precise control over mass transport, reduced sample consumption, multiplexing, and automation; exacerbates sample-driven cross-reactivity through high surface-to-volume ratios, and different substrate materials composing integrated device architectures. As a result, surface chemistry is the pivotal material engineering challenge in biosensor design that determines whether the sensor can function in native biological matrices.

However, implementing reliable surface chemistry passivation strategies compatible with microfluidics is highly nontrivial and they currently fall into two major categories, neither of which addresses this challenge.[5] Protein blocking layers, such as bovine serum albumin or casein, are simple and inexpensive but lack reproducible homogenous surface coverage and still exhibit cross-reactivity.[6] Covalent polymer brushes, like polyethylene glycol (PEG) and PEG-based copolymers, provide better passivation, but their functionalization schemes are often incompatible with microfluidic device fabrication or require multiple sequential layering steps.[7,8] Emerging zwitterionic and hydrophilic polymers are promising alternatives but require complex, multistep grafting procedures.[9] Critically, none of these approaches simultaneously achieves robust surface passivation, chemical tunability, site-specific probe orientation, and substrate compatibility in a single, scalable step. As a result, biosensor performance remains highly dependent on the quality of the input sample, which is typically compensated by upstream processing such as purification, enrichment, or isolation. In the best of cases, such upstream processes simply introduce assay complexity in the form of additional time, labor, or instrumentation demands, and in the worst case, risk introducing sample biases or artifacts in downstream analysis.

Biomimetic lipid coatings, such as supported lipid bilayers (SLBs), represent a promising alternative and have been widely investigated in micro- and nanofluidic devices due to their in situ formation, antifouling properties, and tunable physicochemical characteristics through lipid composition.[10,11] SLBs can further provide stable and versatile surface functionalization, making them attractive building blocks for biosensing applications.[12–14] Nevertheless, converting SLBs into functional biosensing interfaces typically requires multiple sequential in-channel steps for bilayer formation, passivation, and probe functionalization, increasing assay time and introducing variability.[15] To address this materials engineering gap, we developed a one-step in-situ surface functionalization scheme in which engineered nanocarriers rapidly self-assemble into immunoassay-ready lipopolymer coatings within minutes. These lipopolymer coatings follow a design principle that integrates the tuneable biomimetic antifouling properties of lipid membranes, the enhanced passivation of polymer brushes, and the molecular specificity of site-specific oriented capture probes**.** This bottom-up approach decouples surface functionalization from microfluidic device fabrication and establishes a new paradigm for immunoassay surface engineering that proceeds without covalent surface modification.

To validate the materials performance and functional capability of this approach, we apply it to a highly relevant and demanding biosensing challenge in the field of extracellular vesicles (EV), namely to study and characterize EV secretion dynamics directly from unprocessed samples, i.e. in their native environments. [16–23] We chose EVs as they represent an ideal benchmark system, they are inherently heterogeneous, present at low concentrations, and embedded within highly complex biological matrices,[24–26] which render conventional and state-of-the-art digital characterization approaches reliant on upstream isolation steps. These upstream processes bias downstream analysis whether through selective enrichment by size or by molecular composition and restrict assays to end-point measurements due to large sample volume requirements, thereby losing all temporal information associated to EV secretion dynamics.[27] Combining this surface-engineering strategy together with a label-free optofluidic platform, we perform digital immunoassays that time resolve tetraspanin-positive EV secretion dynamics from unprocessed samples from the same population of cells using minimal volumes, paving the way towards continuous real-time analyte monitoring and longitudinal studies. In doing so, we demonstrate that addressing this fundamental biosensor bottleneck does not just improve assay performance, it opens an entirely new experimental regime for studying dynamic biological processes that are inaccessible to existing approaches.

## 2. Results and Discussion

### 2.1 Biomimetic and Rationally Designed One-step Surface Functionalization

Supported lipid bilayers (SLBs) have been widely used as biomimetic lipid coatings for micro- and nano-fluidic devices thanks to their in-situ formation, anti-fouling and tunable physicochemical properties via lipid composition.[9,11,28] In particular, SLBs can withstand high flow rates (>50 µL/min) and support stable, versatile, and robust surface functionalization schemes making them ideal building blocks for biosensing.[10] Nevertheless, conjugating capture probes (antibodies, aptamers, or peptides) to SLBs typically requires multi-step, in-channel workflows that are slow, variable, and difficult to standardize across labs.[15]

Here, we overcome these limitations by engineering modular lipid nanocarriers that integrate all functional building blocks, i.e., passivation, probe conjugation, density of capture sites, and phase state of the final SLB, into a single self-assembling structure. Instead of treating the liposome merely as a nanocarrier, we design its composition to encode the desired properties of the final surface via one-step self-assembly. We used 30 nm sized liposomes composed of three zwitterionic species: POPC, DSPE-PEG2000 and DSPE-PEG5000-azide as the lipid nanocarriers, but other lipid-based assemblies are equally compatible. From this lipid composition, POPC provides a fluid, biomimetic matrix, DSPE-PEG2000 tunes the degree of passivation; and DSPE-PEG5000-azide controls the density of addressable sites for capture-probe conjugation (**Figure** 1a). Together, these components form a rationally engineered nanoscale scaffold that rapidly self-assembles into an immunoassay-ready lipid coating upon contact with the substrate.

Capture probes in the form of IgG antibodies are conjugated prior to SLB formation via biorthogonal DBCO-azide click chemistry. We used Light-Activated Site-Specific Conjugation (LASIC) to covalently crosslink the azide-moiety of the liposome to the antibody upon UV irradiation via a small adapter protein (8 kDa) containing a DBCO moiety with specific affinity to the Fc region of IgG antibodies.[29] This labelling strategy yields site-specific, orientation-controlled antibodies embedded within the liposome corona. Antibody conjugation was confirmed by three orthogonal approaches: transmission electron microscopy (TEM) imaging, dynamic light scattering (DLS) and SDS-PAGE. Specifically, labelled liposomes showed increased specular features in TEM images (Figure 1b), increased hydrodynamic diameters in DLS (Figure 1c), and an upwards shift of the heavy chain in SDS-PAGE confirming the covalent attachment of the small adapter protein to the antibody heavy chain (Figure 1d). The unilamellar morphology of the liposomes was further confirmed by negative staining and Cryo-TEM (SI Figure 1).

We in-situ verify the fast self-assembly of SLBs within microfluidic channels via inline holography[30], by imaging the substrate surface during liposome injection (Supplementary video 1). Although self-assembly of the nanocarriers in contact with the surface suffices to form laterally mobile supported lipid layers, systematic addition of a fusogenic α-helical peptide [31] ruptures any residual intact vesicles on the surface, leading to an optically uniform, passivating and immunoassay-ready surface.

Critically, because the nanocarriers are designed with pre-defined stoichiometry, adjusting the lipid composition prior to liposome formation allows precise and independent control over both the short chain PEG density and capture-probe density (SI Figure 2). The longer polymer chain, PEG5000, associated with the azide group improves immunocapture efficiency by reducing steric hindrance and improving target accessibility. For instance, 0.3 mol% DSPE-PEG5000-azide yields an estimated incorporated capture-probe density of one molecule per $15 \times 15$ nm$^2$, corresponding to an antibody surface coverage below 50%, which has been found to be ideal for EV capture.[32] Although SLB formation was also observed with up to 1 mol% DSPE-PEG5000-azide, we found that the 0.3 mol% captured more EVs, consistent with reduced steric crowding at lower capture probe density and with previous quantification of capture-moiety density using biotin-Neutravidin systems.[31,33] Nevertheless, the capture probe density can be easily adjusted and optimized for other targets by changing the doping percentage.

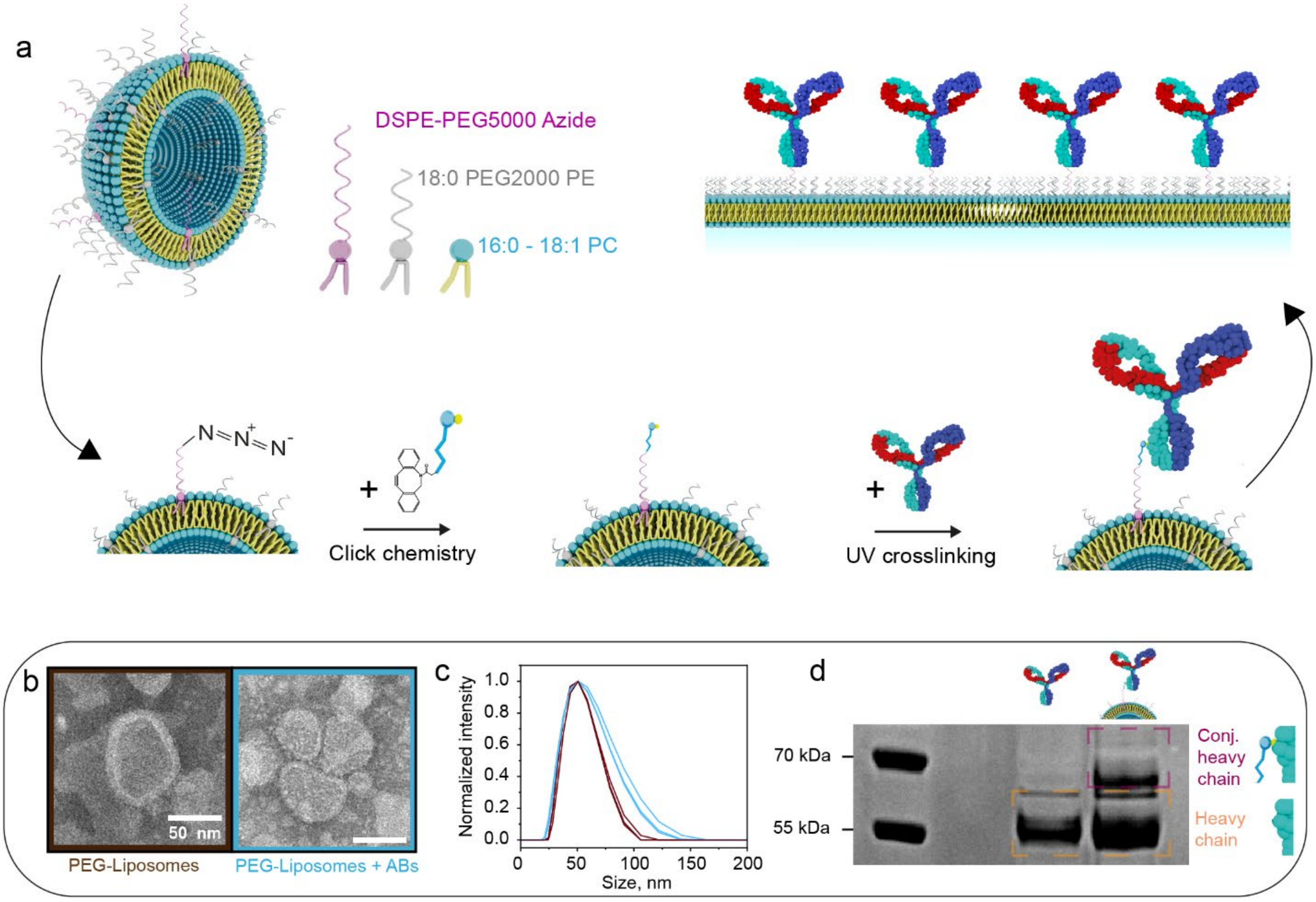

**Figure 1 | Design and characterization of biomimetic PEGylated liposomes for surface passivation and functionalization.** (a) Schematic of the lipid nanocarriers comprising PEGylated liposomes composed of POPC, DSPE-PEG2000, and DSPE-PEG5000-azide lipids. Antibodies are site-specifically and covalently labelled to the liposomes using a Light-Activated Site-Specific Conjugation strategy: a bio-orthogonal click reaction attaches a small adapter protein with specific affinity to the Fc-region of IgG antibodies to the azide groups on the liposome, followed by UV-crosslinking. When introduced into a microfluidic channel, the liposomes self-assemble into a functional lipid coating, enabling downstream immunoassays. For clarity the schematic excludes any antibodies on the proximal (substrate-facing) leaflet. (b) Representative TEM images of liposomes before and after antibody conjugation. (c) Dynamic light scattering measurements show an increase in hydrodynamic diameter upon conjugation . Three independent measurement curves are shown. Scale bars: 50 nm (d) SDS–PAGE analysis confirms successful conjugation, with a characteristic upward shift of the antibody heavy chain.

## 2.2 Hybrid PEG Polymer Brush Lipid Coatings Minimize Sample-driven Cross-reactivity

Having established the one-step surface functionalization scheme, we optimized the lipid composition to minimize sample-driven cross-reactivity. Specifically, we tuned the surface density of the short-chain PEG polymer moiety, known to passivate against complex environment,[34] by varying the molar composition of DSPE-PEG2000. Figure 2 summarizes the results of the systematic variation of PEG polymer surface density across different conformational grafting regimes determined according to P.G. de Gennes' grafting rules (SI Figure 3).[35] Specifically the mushroom, transition and polymer brush regimes were produced at 1%, 5% an 10% molar doping of DSPE-PEG2000, in good agreement with literature.[34] The passivation from the different hybrid lipid coatings across grafting regimes was then assessed by challenging the surface with a panel of high-concentration and compositionally diverse samples: 100 nm sized POPC liposomes ($10^{12}$ particles/ml), H358 derived EVs (5x $10^{10}$ particles /ml), 1 mol% cationic DOTAP doped 100 nm POPC-liposomes ($10^{12}$ particles/ml), as well as culture medium (RPMI1640) supplemented with 10% or 50% fetal bovine serum (FBS) or 100% FBS. Figure 2b shows that as the DPSE-PEG2000 doping percentage increased and approached the polymer brush regime, the sample-driven cross-reactivity was minimized across all sample types (Supplementary video 2). Even against undiluted FBS, hybrid lipid coatings with 10% molar doping of DPSE-PEG2000 showed no detectable binding events (Figure 2a, Supplementary video 3). In summary, the final optimized surface minimizes sample-driven cross-reactivity against highly concentrated neutrally (POPC liposomes), positively (DOTAP doped liposomes) and negatively (EV) net-charged particles as verified by zeta-potential measurements (SI Figure 4), as well as against complex protein mixtures in culture medium and FBS.

Finally, we assessed the lateral mobility of the lipopolymer coatings as well as their passivation performance on PDMS. For lipopolymer coatings formed on glass, FRAP measurements showed almost full fluorescence recovery (mobile fraction > 0.97) across nanocarrier formulations containing 0, 1, 5 and 10 mol% PEG2000 doping, and for the 10 mol% PEG2000 formulation after antibody conjugation and fusogenic peptide treatment (SI Figure 5). In contrast, the coating on PDMS, prepared at 10 mol% PEG2000 doping, showed a lower mobile fraction (0.74, SI Figure S6). Despite this lower mobility, the PDMS coating still passivated the surface with similar performance to a glass one when challenged with EVs at $5\times10^{10}$ particles/ml, whereas untreated PDMS showed pronounced nonspecific EV binding (SI Figure 6). Together, these results show that self-assembled lipopolymer coatings extend across the entire microfluidic chamber, including both glass and PDMS substrates, and that functional passivation does not require high lateral mobility, in agreement with recent related works. [36]

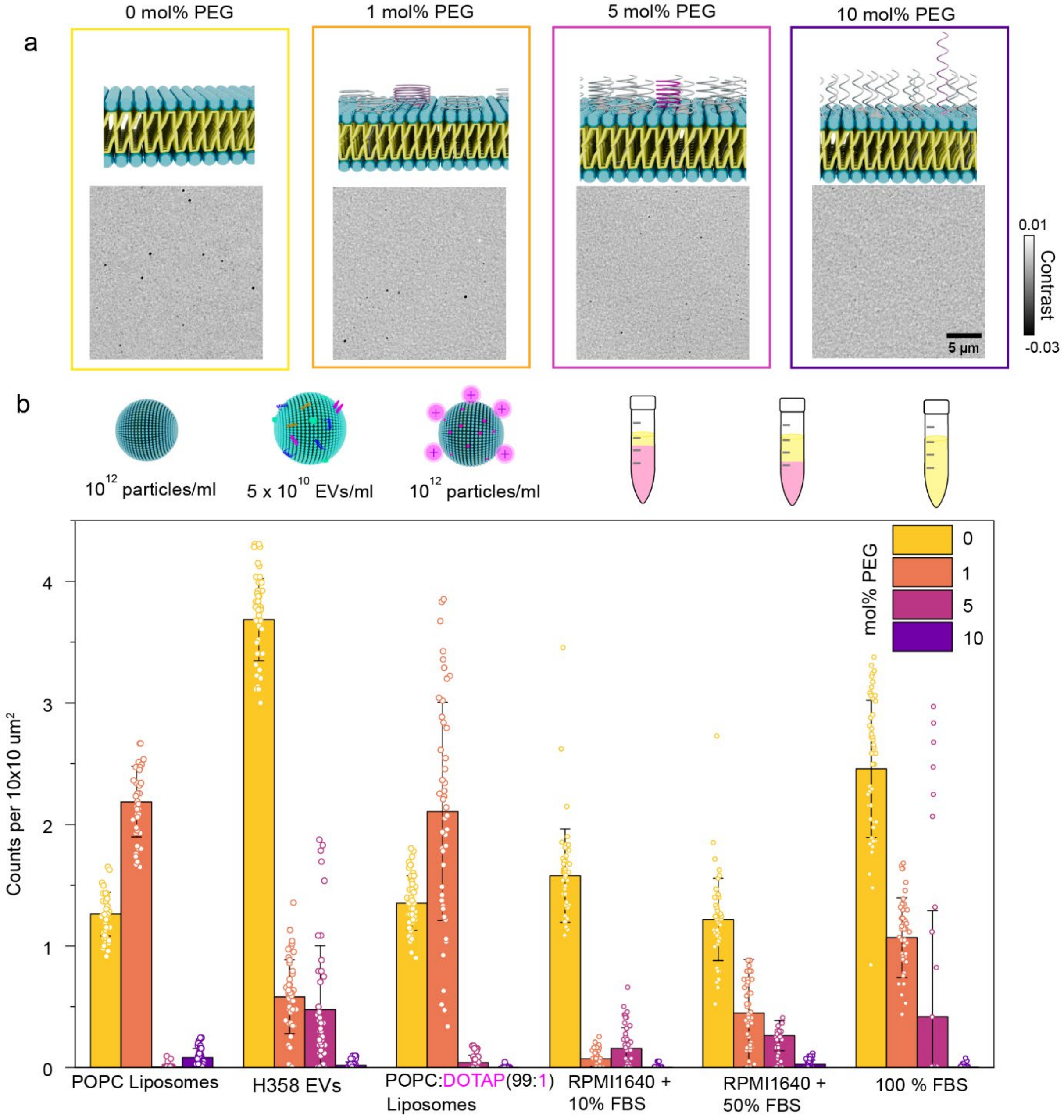

**Figure 2 | Tuning PEGylated lipid density to optimize bilayer passivation.** (a) Top: conceptual illustration of the conformation transition from the mushroom to the polymer brush regime in the biomimetic lipid coatings upon increasing the PEG2000-lipid surface density. Bottom: representative label-free images of each lipid coating measured against full serum (FBS). As the PEG-2000 surface density increases and transitions into the polymer brush regime, sample-driven cross-reactivity in the form of non-specific binding reduces. Scale bar: 5 µm (b) Quantitative passivation tests comparing lipid coatings with different PEG surface densities against a panel of test samples, including neutral POPC liposomes, extracellular vesicles (EVs), cationic DOTAP liposomes, and cell culture medium supplemented with 10%, 50%, or 100% FBS. The bar plot represents the mean of detected features per single FOV over an area scan ± SD over N independent FOVs (N=52).

### 2.3 EV-mimetic Liposomes as Test Sample

To validate target-specific capturing and compatibility with label-free digital biosensing assays, we generated biotin-doped EV-mimetic liposomes and formed polymer brush hybrid SLBs functionalized with anti-biotin antibodies (Figure 3a). The EV-mimetic 100 nm sized liposomes were characterized by NTA and DLS (SI Figure 7) and their average biotin density was tuned by adjusting the biotinylated lipid fraction prior to preparation. We then performed two dose response assays by varying either: i) the total liposome concentration or ii) the biotin density per liposome.

The optofluidic platform used for performing the assays features a negative pressure controlled multiplexable microfluidic chip placed on a digital in-line holographic microscope building up on earlier work.[31] In the liposome concentration assay, we fixed the outer biotin density to 50 molecules per liposome (SI Figure 2), representing the typical expression levels of single tetraspanin proteins ubiquitously found in EVs.[37] Figure 3b shows a clear and robust dose response of triplicate measurements, alongside representative zoom-in images of a 10x10 $\mu m^2$ area of the digital biosensor. As a negative control, a surface labelled with non-specific mouse IgG antibody showed negligible non-specific binding. In the biotin density assay, we fixed the liposome concentration at $5x10^9$ particles/ml and observed a clear dose response as the biotin density increased from 0 to 500 molecules per liposome (Figure 3c). Double negative controls, EV-mimetic liposomes with no biotin and surfaces labelled with non-specific mouse IgG antibody, again showed minimal non-specific interactions. Fits to the Hill equation (solid lines) confirmed the expected binding model, the compatibility of our approach with digital immunoassays, and revealed that the sensor responds more strongly to changes in the total number of target antigens per particle ($\eta$ =2.49) than to the total number of antigen-bearing particles ($\eta$ =1.16).

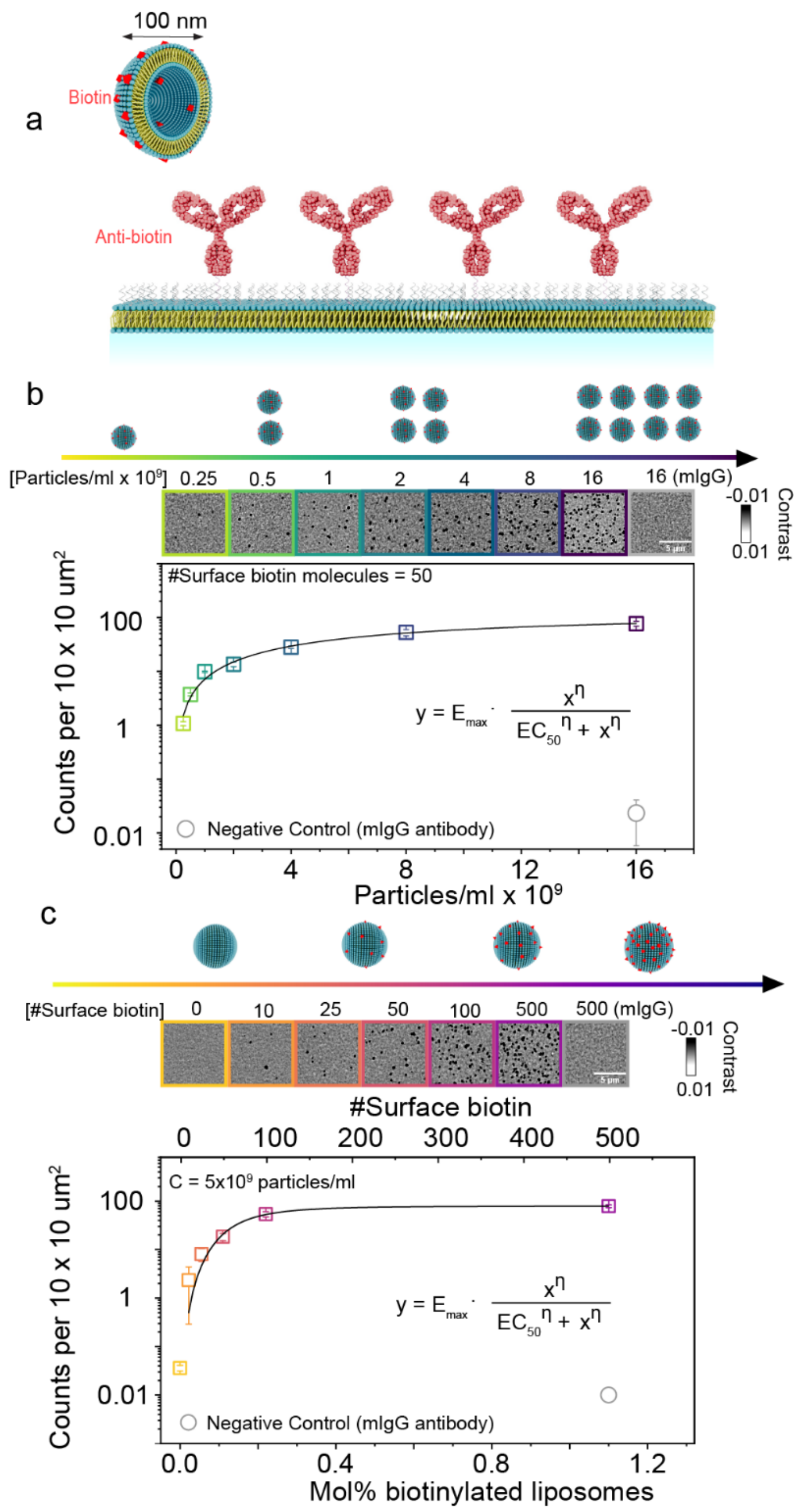

**Figure 3 | Dose–response characterization of EV-mimetic liposomes.** (a) Schematic of the assay concept: liposomes (~100 nm) containing biotinylated lipids are captured by surface-immobilized anti-biotin antibodies, mimicking tetraspanin-based EV sensing. (b) Dose–response curve obtained by varying particle concentration while keeping the fraction of biotinylated lipids constant (N = 3). (c) Complementary dose–response obtained by fixing particle concentration and varying the amount of incorporated biotinylated lipids Representative zoom-in of scans corresponding to the plotted data are shown above each curve. Data points represent mean ± SD over 3 independent chip replicas (N=3). Scale bars: 5 µm.

## 2.4 Time-resolved Tetraspanin Profiling of EVs Directly from Cell Culture Supernatant

Tetraspanins are key constituents of EVs and participate in many critical processes such as endosomal sorting, cellular targeting in cell-cell communication and defining cellular identity.[38,39] Leveraging our developed surface functionalization scheme and label-free optofluidic platform capable of digital immunoassays, we directly profiled tetraspanin-positive EVs from unprocessed human bronchioalveolar carcinoma cell supernatants (H358). The fast one-step functionalization together with low sample consumption from each microfluidic channel enabled time-resolved profiling from the same cell culture, including technical triplicates to account for chip-to-chip variability. For each time point, 600 µl (3 replicates, 4 markers, 50 µl aliquot) of H358 supernatant was sampled from a T-150 flask and immediately replaced with fresh medium to keep the total cell culture volume of 20 mL constant (Figure 4a).

We specifically quantified the expression level of three pan-EV tetraspanin markers CD9, CD81 and CD63, together with a negative control (mIgG), from repeatedly drawn aliquots from the same cell culture over 36 hours in 12-hour intervals (Figure 4b). To compare the relative expression level between the different markers independent of the total EV concentration, we normalized the digital particle counts. The relative tetraspanin expression profile trends were highly reproducible and consistent for all time points: CD9 >> CD81 > CD63 > mIgG (Figure 4c). Comparison with 72-hr endpoints measurements from supernatants from a different cell culture flask and cross-validation from EVs purified via size-exclusion chromatography (SEC) and characterized via NTA, Western Blots (WB) and TEM imaging (SI Figure 9) showed no statistically significant differences (one way ANOVA, $p > 0.05$). These results demonstrate that our platform can directly quantify and monitor the secretion dynamics of tetraspanin-positive EV subpopulations from the same ensemble of cells without the need for time-consuming, lossy, and potentially bias-introducing isolation steps, thereby delivering capabilities and accessing information inaccessible to standard isolation workflows.

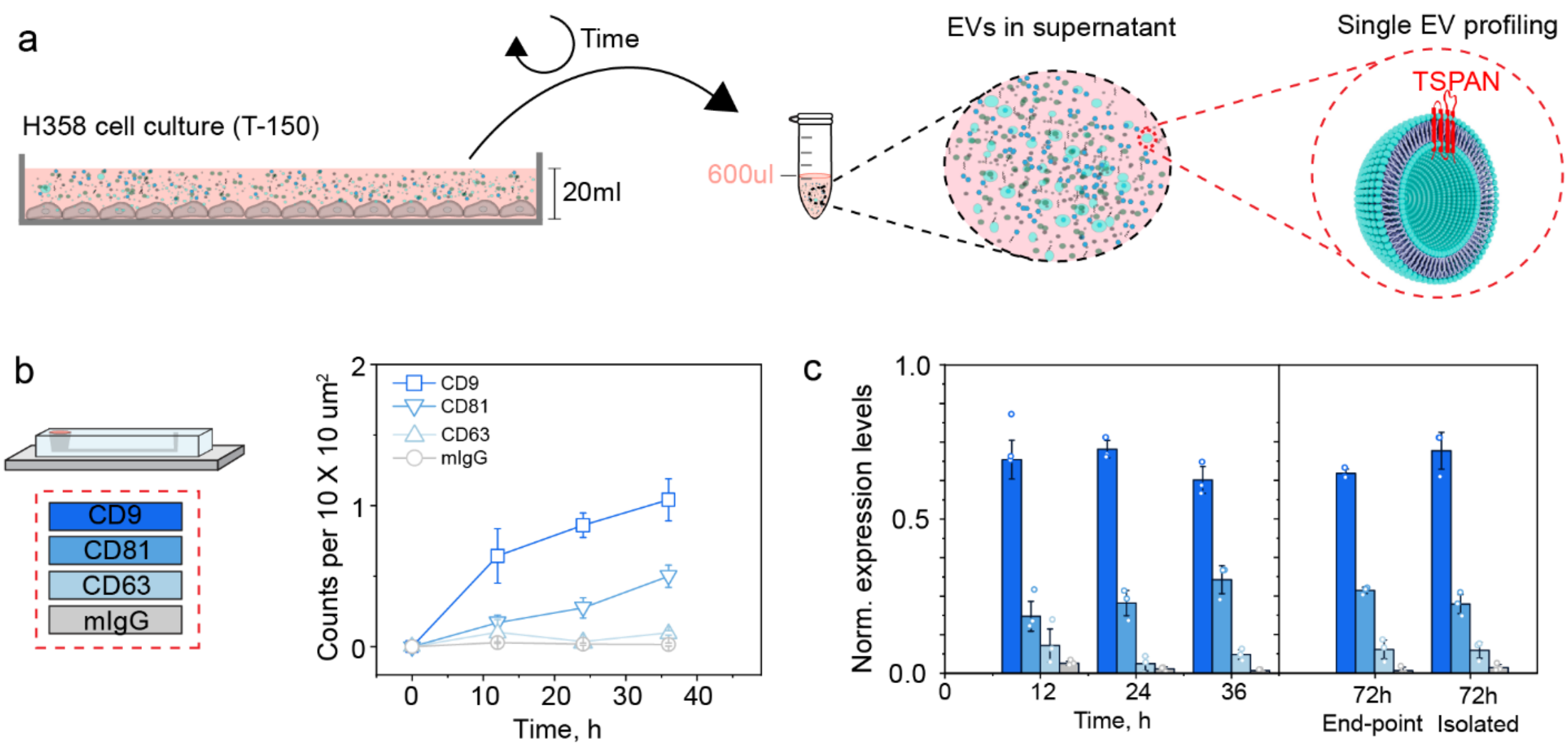

**Figure 4 | Time-dependent profiling of EV markers in H358 2D culture.** (a) Schematic of time-resolved pan-EV marker tracking in H358 cells cultured in T-150 flasks. (b) Normalized counts (per 10 × 10 µm² scanned area) for CD9, CD81, CD63, and mouse IgG (negative control) at multiple time points (N = 3). (c) Normalized count density plots of the time-dependent measurements shown in (b) and additional end-point measurements of culture supernatant from an independent flask and EVs isolated by size-exclusion chromatography, showing comparable tetraspanin expression profiles. No statistically significant differences for the individual biomarkers were observed across datasets (one-way ANOVA, $p > 0.05$). Data is represented as mean ± SD of 3 independent channel replicas of which one replica is within the same chip and one in a different chip (N=3).

### 2.5 Continuous EV Concentration and Secretion Rate Estimation

The secretion rate of EVs is a straightforward metric that reports on the current state of the parent cells without requiring knowledge of the cargo trafficked, and has so far been applied to track physiological and pathological changes.[40,41] Nevertheless, assessing the secretion rate directly from supernatants is challenging due to the interferences introduced by the complex matrix they are in,[42] or by the difficult-to-quantify losses introduced by upstream sample processing workflows. Reported secretion rates therefore vary widely with cell type, cell state, and culture conditions. For example, Auber et al. showed secretion rates ranging from 0.003 to 45 EVs/cell/min depending on the originating cell line.[43] Whereas, Chiu et al. showed secretion rate variations depending on drug-treatment, cell culture medium pH, or growth factor treatment of the parent cells.[44]

To obtain an unbiased secretion rate directly from unprocessed supernatant (Figure 5a), we first generated a calibration curve using SEC-isolated EVs from H358 with known concentrations determined by Nanoparticle Tracking Analysis (NTA, Figure 5b). This calibration is necessary as it captures the specific kinetics of each antibody-antigen pair, as well as the dependence on the average number of target antigens per particle (Figure 3c). We selected CD9 as the capture target due to its highest abundance in H358-derived EVs compared to other pan-EV tetraspanins (Figure 4c), providing the largest dynamic range. Across the tested concentration range, the CD9-specific digital immunoassay exhibited a highly linear response ($R^2$ = 0.997) with a calculated LOD of 2.2 x $10^8$ EVs/ml.[45] Importantly, this calibration is applicable to direct supernatant measurements because the negative controls of isolated and non-isolated EV supernatants remained negligible at all time points, and EVs were derived from the same cell line cultured under identical conditions (Figure 4c).

We next sampled 300 µL of H358 culture supernatant twice daily (3 technical replicates, 2 markers, 50 µl aliquot) replacing each aliquot with fresh medium. EV secretion rate was then determined from correlative cell count and EV concentration measurements over several time points (Figure 5c). Cell numbers were quantified from brightfield microscope images using a machine-learning segmentation model, while the concentration of CD9-positive EVs were determined from the digital immunoassay using the above calibration (Figure 5d). At each time point a negative control using mIgG was performed showing minimal sample-driven cross-reactivity. Linear fits to both cell and EV counts (both $R^2 = 0.97$) yielded a global secretion rate of 2.78 EVs·cell$^{-1}$·min$^{-1}$. Crucially, this value is measured directly from the unprocessed sample, free from isolation losses or purification biases and represents a better estimate of the EV secretion dynamics.

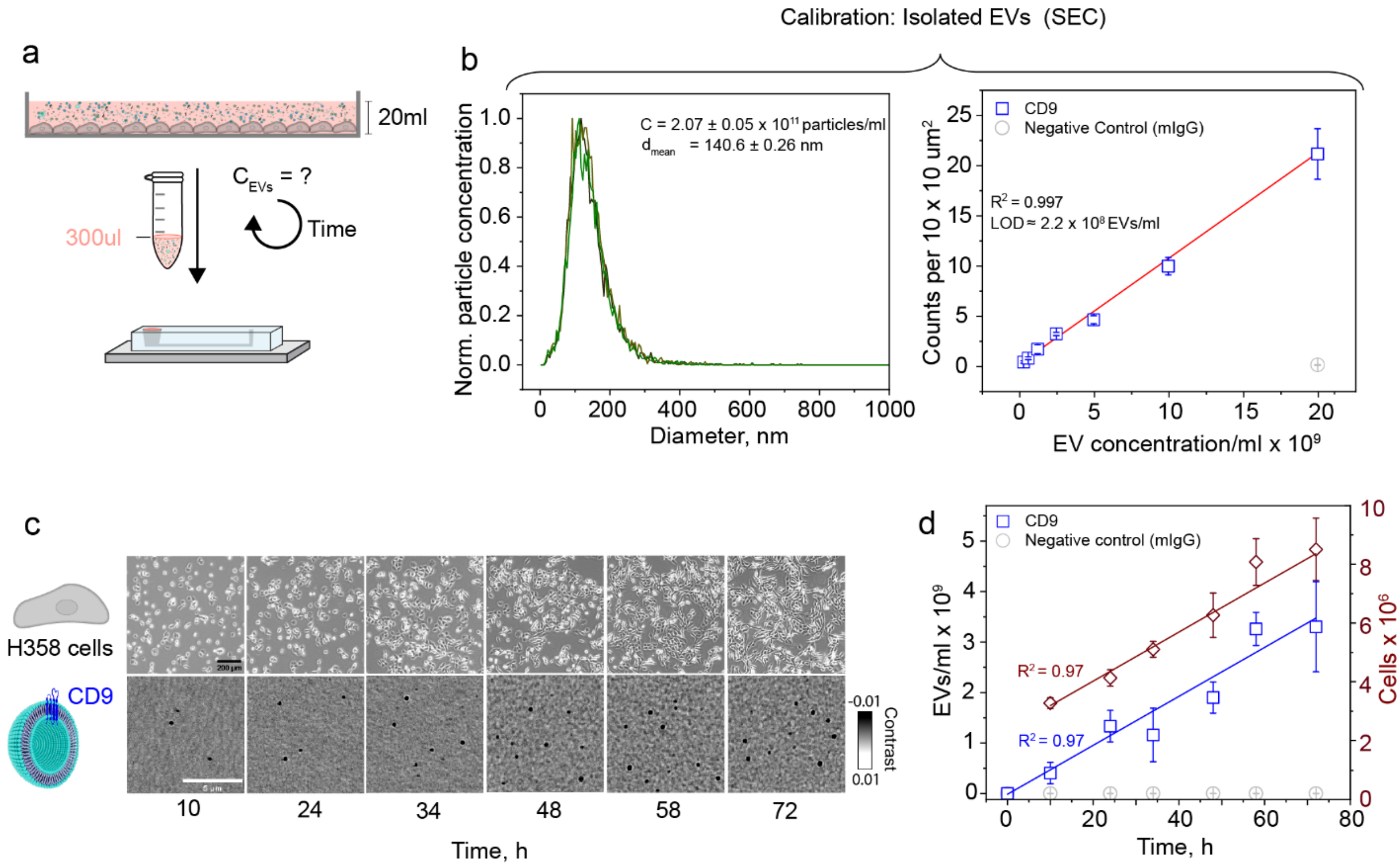

**Figure 5 | EV concentration estimation of H358 2D cultures**. (a) Schematic of EV concentration assessment over time in H358 cells cultured in a T-150 flask. (b) Sensor calibration using EVs isolated by size-exclusion chromatography (SEC): a stock concentration was determined by NTA, and a dilution series produced a linear response ($R^2 = 0.997$). For NTA, the data is shown as three curves of three independent measurements, values are stated as mean ± SD. The dilution series is represented as three independent channel replicas in three different chips (N=3). (c) Inverted microscope images of H358 cells over time, paired with representative zoom-ins of scans of immobilized CD9-positive EVs captured directly from supernatant. Scale bars: 200 µm and 5 µm. (d) The calibration curve from (b) was applied to time-resolved supernatant measurements. In combination with cell counts obtained via machine learning-based analysis, this allows calculation of a global EV secretion rate using linear fits to both curves. Cell count data is represented as mean ± SD of 5 different FoV. EV count data is represented as

3 independent channel replicas of which one replica is within the same chip and one in a different chip (N=3).

## 2.6 Dynamic EV Secretion Analysis from Low-volume Tissue-like Spheroids

To demonstrate the ability of our functionalization scheme paired with a label-free digital immunoassay to handle limited sample volumes and low cell numbers that are incompatible with upstream purification steps, we applied our approach to a more physiologically relevant three-dimensional spheroid cell model. For that, 30’000 H358 cells were seeded in InSphero culture plates to generate uniform spheroids. To avoid perturbing the system with frequent supernatant sampling, we took advantage of the recommended spheroid culture protocol in which culture medium is exchanged daily. The collected media after exchange, which under normal conditions are discarded, were pooled and analyzed for EV secretion over a 10-day period. Specifically, we profiled the supernatant for the pan-EV markers CD9, CD81, and CD63, while assessing the sample-driven cross-reactivity with an isotype control (mIgG), similar to the 2D model workflow (Figure 6). Compared to the 2D cell cultures, CD63 expression was reduced in spheroids, while CD9 and CD81 showed comparable profiles across both systems.

To mimic a time-resolved functional measurement on the same 3D system, spheroids were subjected to thermal stress after a four-day maturation period: 39 °C for two days, followed by 41 °C for two days, then returned to 37 °C for recovery. Across all time points, the relative expression profile CD9 >> CD81 >> CD63 ≈ mIgG remained consistent. Notably, elevated temperatures induced a decrease in EV secretion, which became more pronounced at 41 °C. Upon restoring normal conditions (37°C), EV release increased again. Although the underlying biological responses to thermal stress are complex and beyond the scope of this study, these experiments highlight that our surface-functionalization scheme enables longitudinal dynamic, isolation-free tracking of EV secretion directly from supernatant, even in challenging 3D tissue-like environments, which are simply inaccessible to any other approach that relies on upstream sample processing.

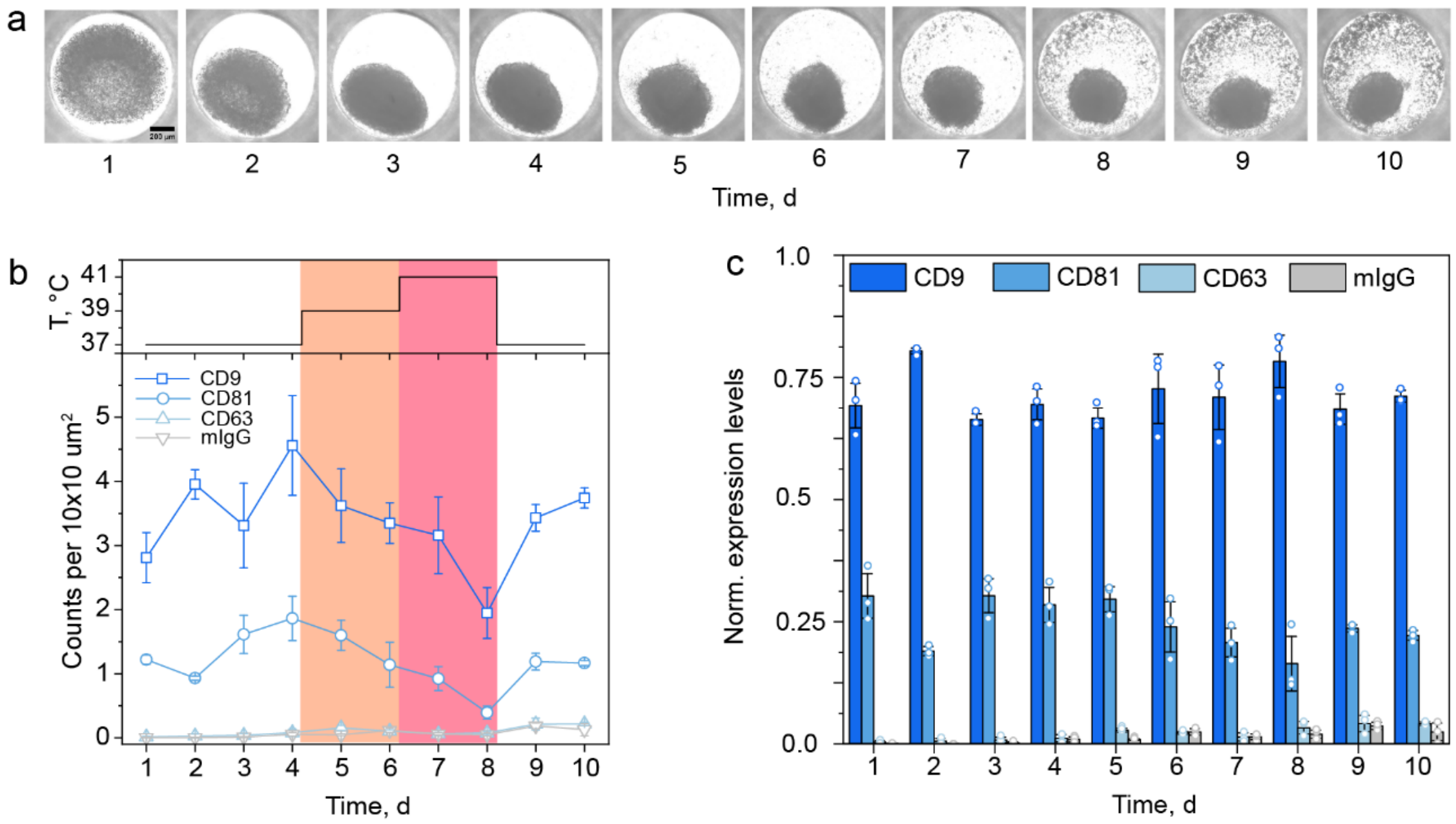

**Figure 6 | Time-dependent EV tracking of H358 spheroids with temperature perturbation.**
(a) Inverted microscope images of lung cancer (H358 cell line) spheroid formation and evolvement over the time course of the experiment (Scale bar: 200 um). (b) Normalized counts (per 10 × 10 μm² scanned area) of the immobilized EVs for CD9, CD81, CD63, and mouse IgG as negative control while the cell system was perturbed by temperature changes to 39°C for days 5 and 6 and 41°C for days 7 and 8 (n=3 for all data points). (c) Corresponding tetraspanin profile to the counts. Data is represented as 3 independent channel replicas of which one replica is within the same chip and one in a different chip (N=3).

## 3. Conclusion

This work establishes a general design principle for functional biosensing interfaces based on engineered lipid-based nanocarriers that spontaneously assemble into antifouling, immunoassay-ready lipopolymer coatings in a single step that are compatible with microfluidic-based sensor architectures. This design principle combines precise control over the PEG density and bio-orthogonal chemistry-based probe incorporation to form tuneable “plug-and-play” surface coatings, which are decoupled from device fabrication, and simultaneously suppress sample-driven cross-reactivity whilst delivering target specificity. Because the nanocarriers self-assemble on both glass and PDMS, a single injection coats and passivates the entire microfluidic channel. By suppressing sample-driven reactivity, this approach enables direct detection of analytes in their native biological environments, thereby removing potential measurement biases introduced by isolation and purification workflows that have shaped biosensing practice for decades. When combined with the advantages of low-volume sample handling from microfluidics and ultrasensitive label-free microscopy, this approach is compatible with real-time digital detection of analytes such as EVs embedded within their complex native sample matrices.

Beyond eliminating the requirement of upstream sample processing steps, the significance of our surface engineering approach lies in the experiments it makes feasible. Its preservation of native sample conditions makes this approach well suited for systems where quality control is paramount, such as therapeutic EV manufacturing[46,47], where the potential effects of sample processing-induced artifacts remain poorly characterised. Similarly, the combination of low-volume sample handling with the isolation-free performance of these lipopolymer coatings unlocks time-resolved measurements regimes inaccessible to conventional static end-point workflows. Applied here to longitudinal monitoring of EV secretion from 2D and 3D lung cancer models, this capability reveals secretion dynamics and molecular signature shifts that endpoint, isolation-dependent assays fundamentally cannot probe. In doing so, we establish a path towards a new class of functional, time-resolved EV assays that can elucidate dynamic biological processes previously masked by sample-volume limitations and purification workflows, such as probing the dynamic responses of cellular state transitions and stimulus-dependent EV secretion behaviours.

As such we believe that by tackling a fundamental biosensor bottleneck, our work offers a generalizable route for advancing immunoassay surface engineering beyond current detection approaches and enabling the detection of target molecules from challenging low-volume samples, such as organ-on-a-chip systems. In summary, these results establish a new paradigm for integrating biomimetic surface engineering into biosensors, bridging a critical gap between sensitive molecular sensing and practical yet reproducible implementation under native biological conditions.

## 4. Experimental Section

*Microfluidic System:* Microfluidic chips were fabricated using a wafer-based soft lithography process. SU-8 GM 1070 negative photoresist (Gersteltec Sàrl) was spin-coated onto a silicon wafer to reach a thickness of 100 µm and patterned via direct laser writing (Heidelberg µMLA, 365 nm) based on a custom 2D AutoCAD design. Following development, the wafer surface was rendered hydrophobic by vapor-phase silanization using chlorotrimethylsilane (Thermo Scientific Chemicals, 110122500) in a vacuum chamber. Polydimethylsiloxane (PDMS; Sylgard 184, Dow Corning) was prepared at a 10:1 base-to-curing-agent ratio, poured over the patterned master, and cured at 80 °C for 1 h in a convection oven. The cured PDMS layer was then peeled off, trimmed to size, and inlet reservoirs and outlet holes were punched. Glass cover slips ($24 \times 50$ mm$^2$, thickness $0.17 \pm 0.01$ mm; Karl Hecht) were used as substrates and irreversibly bonded to the PDMS by oxygen plasma activation (Diener Electronic, Atto, 13.56 MHz, 300 W, 1.5 sccm) for 30 s. Assembled devices were post-baked at 80 °C overnight to enhance bonding stability and fully cure the PDMS. For flow generation, a multi-channel peristaltic pump (Reglo ICC, Ismatec) was used. Negative pressure was applied at the channel outlets, while fluid was introduced into the inlet reservoirs via manual pipetting. The resulting flow rates were set by the peristaltic pump settings. The microfluidic chip consists of 8 parallel channels allowing multiplexed functionalization of several channels at the same time (SI Figure 10). All measurements were performed as N=3, corresponding to three different chips replicas or one in chip and one chip to chip replica for measurements where only four channels were needed.

*Surface Functionalization:* Engineered lipid nanocarriers in the form of liposomes consisted of 1-palmitoyl-2-oleoyl-sn-glycero-3-phosphocholine (16:0-18:1 PC) (850457 C, Avanti Polar Lipids, Inc), 1,2-distearoyl-sn-glycero-3-phosphoethanolamine-N-[methoxy(polyethylene glycol)-2000] (18:0 PEG2000 PE) (880120 C, Avanti Polar Lipids, Inc) and 1,2-distearoyl-sn-glycero-3-phosphoethanolamine-N-[azido(polyethylene glycol)-5000] (DSPE-PEG(5000) Azide) (880227P, Avanti Polar Lipids, Inc) at a molar ratio of 89.7:10:0.3 if not stated otherwise. All lipids were mixed in chloroform and dried overnight under vacuum. Following rehydration with filtered (0.2 µm) PBS buffer (10010023, Thermo Fisher Scientific) to a concentration of 5 mg/ml and 1 ml volume, the lipids were extruded to form functional liposomes. For that, the lipid stock was passed sequentially 21 times through polycarbonate membranes with pore sizes of 100nm, 50nm and 30nm at an extruder temperature of 65°C. The extruded liposomes were diluted to a concentration of 1 mg/ml and incubated with oYo-Link® mIgG1 DBCO (AT3003- mIgG1, AlphaThera, Inc.) at 4°C overnight with a 1.5 molar excess of DBCO to the liposome incorporated azide. Desired antibodies were then crosslinked to the liposomes according to manufacturer's instructions at a 1:10 ratio of antibody to oYo-Link® to ensure near complete conjugation of all antibodies by exposure to UV light (365 nm, LED Photo-Crosslinking Device (LEDPX), AlphaThera, Inc.) for 2h. Conjugated Antibodies to the liposomes were anti-CD63 (mAB, preservative free, ANC-215-820, Ancell), anti-CD81 (mAB, preservative free, ANC-302-820, Ancell), anti-CD9 (mAB, preservative free, ANC-156-820, Ancell), mouse IgG1 Isotype control (mAB, preservative free, AG-35B-0003PF, Adipogen) or anti-biotin (mAB, Azide Free Clone B-E24, 857.060.005., Medix Biochemica). The final antibody-functionalized liposomes were diluted to a concentration of 0.1 mg/ml before injection to the microfluidic channels. The conjugated liposomes were stored at 4°C and used within 5 days. The size of the liposomes was verified with dynamic light scattering measurements (DLS, Malvern Zetasizer Nanoseries Nano-ZS, $T = 25$ °C, 173 backscatter detection).

*Lipid Nanocarrier Characterization:* Successful conjugation of the antibodies to the oYo-link molecule was verified with SDS-page**.** For that a total of 2 µg per lane was loaded for three conditions. One lane with the oYo-link reagent only, one with the antibody only, and one with the oYo-link conjugated to the liposomes and crosslinked to the antibody. All samples were mixed with 10% Bolt Sample Reducing Agent (B0009, Invitrogen) and 25% LDS Sample Buffer (B0007, Invitrogen). Samples were resolved on a Bolt 4–12% Bis-Tris Plus WedgeWell gel (NW0412120BOX, Invitrogen). The gel was run at 200 V for 36 min in a Mini Gel Tank (A25977, Invitrogen) using 1× Bolt MES Running Buffer (B0002, Invitrogen). After electrophoresis, the gel was stained with SimplyBlue SafeStain (LC6060, Invitrogen) for 1 h on a rocker at room temperature, destained in dd$H_2O$ for 2h and imaged using iBright CL1500 instrument (A44240, Thermo Scientific).

Liposome morphology and shape was accessed by negative staining TEM as well as Cryo-TEM imaging: For TEM visualization aliquots of 3 µl of the dispersion were placed on glow discharged (Emitech K100X, GB) or untreated carbon-coated grids (Quantifoil, D) for 2 min. The excess of fluid was removed with filter paper and the grids were washed two times with distilled water followed by draining off excess liquid with filter paper. The wet samples were stained with a drop of 2% phosphotungstic acid (pH 7.3) for 1 s, followed by a second step of 20s. Excess moisture was drained off with filter paper, and the imaging of the air-dried grids was done with a TEM Morgagni 268 (Thermo Fisher Scientific (TFS), USA) operated at 100 kV in bright field mode.

For cryo-TEM, the samples were prepared in a controlled-environment vitrification system Vitrobot Mark IV (TFS, USA) at 22 °C and 100 % humidity. Next, 3.6 µL of sample was placed on hydrophilized lacey carbon-coated copper grids (EMS, USA). To form a thin film on the mesh holes, excess sample was blotted away with filter paper. Then, the grids were plunge frozen into a mixture of liquid ethane/propane cooled by liquid nitrogen. For automated loading in the cryo-TEM, the vitrified grids were clipped into AutoGrid sample carriers (Thermo Fisher Scientific (TFS, USA). Data were collected with a Titan Krios (TFS, USA) operated at 300kV of acceleration voltage equipped with a Falcon III direct electron detector (TFS, USA). Digital images were recorded under low dose conditions and an underfocus of 1-3 µm to get enough contrast.

*EV-mimetic liposomes and EV samples:* For the test system of EV-mimetic nanoparticles, liposomes were extruded as described above through a polycarbonate membrane with a pore size of 100nm. The composition of the Liposomes consisted of 1-palmitoyl-2-oleoyl-sn-glycero-3-phosphocholine (16:0-18:1 PC) (850457 C, Avanti Polar Lipids, Inc) with incorporated 1,2-dioleoyl-sn-glycero-3-phosphoethanolamine-N-(cap biotinyl) (18:1 Biotinyl Cap PE) (870273 C, Avanti Polar Lipids, Inc) according to molar ratios shown in Figure 3. Particle concentration and size distribution was determined with NTA (ZetaView PMX 120-Z, Particle Metrix GmbH) and DLS (Malvern Zetasizer Nanoseries Nano-ZS, $T = 25$ °C, 173 backscatter detection) measurements.

EV samples were either purified and isolated from human bronchioalveolar carcinoma H358 (ATCC-CRL-5807, passage number <15) cell culture supernatant or supernatant taken directly from culture flasks or culture plates. For the isolated EVs and real time tracking of 2D cell models, cells were seeded in T-150 culture flasks. Spheroid formation was achieved with InSphero culture plates (Akura™ 96 Spheroid Microplates) with V shaped and low attachment coated wells. All cells were cultured in a humidified incubator at 37°C and 5% $CO_2$. For end-point measurements, the supernatant was harvested after 72h of culture at a cell confluency of approximately 60-70%. Culture medium consisted of 90% RPMI 1640 (21875-034, Thermo Fisher Scientific) supplemented with 10% EV-depleted FBS (A2720801, Thermo Fisher Scientific).

For measurements with purified EVs, EVs were purified and concentrated with size-exclusion-chromatography (SEC) columns (Izon, qEV1 Gen2,70nm) according to manufacturer's instructions. In short, the supernatant was centrifuged at 1500g for 10min and transferred to new tubes, following centrifugation at 10'000g for another 10min. Supernatant was then concentrated with 50kDa molecular weight cut-off filters (Amicon® Ultra Centrifugal Filter, UFC905096, Sigma-Aldrich) before passing through the SEC columns. Fractions 1-4 were collected and concentrated again with the molecular weight cut-off filters. Protease inhibitor (78440, Thermo Fisher Scientific) was added to the purified and isolated EV samples and the samples were stored at -80°C.

For unpurified EV assays, the samples were taken directly from the supernatant and were stored at 4°C after addition of protease inhibitor and measured either the same or the next day.

*EV characterization:* EV concentration and size distribution of isolated EV samples were determined by NTA measurements (ZetaView PMX 120-Z, Particle Metrix GmbH). The successful isolation of EVs was validated by TEM-imaging and Western blot (WB) analysis. TEM imaging was performed the same way as for the liposomes.

For WB analysis, EV-specific proteins (CD63, CD9 and TSG101) as well as the contamination marker (GRP94) were detected. To obtain EV lysates, EV suspensions were lysed with 10 X RIPA Buffer containing protease inhibitor cocktail (Merck 20-188, Thermo Scientific 87786). For the whole-cell lysate, the cell layer was washed 2x with PBS and then the cells were lysed with 1x RIPA Buffer with protease inhibitor cocktail (Thermo Scientific 89900, Thermo Scientific 87786). Protein concentrations were determined using Pierce BCA Protein Assay Kit (Thermo Scientific 23227). 15 µg of EV sample and whole-cell lysate were mixed with 10% Bolt Sample Reducing Agent (Invitrogen B0009) and 25% LDS Sample Buffer (Invitrogen B0007) and heated up to 72°C and mixed in a thermomixer for 10min. Protein samples were then resolved on a Bolt 4-12% Bis-Tris plus Wedge Well gel (Invitrogen NW0412120). The gel was run at 200V for 28min in a Mini Gel Tank (Invitrogen A25977) using 1x Bolt MES Running Buffer (Invitrogen B0002). Protein transfer onto a nitrocellulose membrane (Invitrogen IB23002) was performed using the iBlot2 western blot transfer device (Invitrogen IB21001) with Program 0 (20V for 1min, 23V for 4min, 25V for 2min). After the transfer, the membrane was washed with ddH20 and blocked with EveryBlot Blocking Buffer (Biorad 12010020) for 1h at room temperature. Primary antibodies were diluted in EveryBlot Blocking Buffer (Biorad 12010020) and membrane was incubated overnight at 4 °C on a rocker. Following three 10 min washes in TBS-T, secondary horseradish peroxidase (HRP)-conjugated antibodies in EveryBlot Blocking Buffer (Biorad 12010020) were added for 2 h at room temperature. Following antibodies and dilutions were used: CD9 (1:500,10626D, Thermo), GRP94 (1:1000, FNab03665, FineTest), TSG101 (1:1000, orb1564135, biorbyt), CD63 (1:1000, M01080-1, Boster), Anti-rabbit IgG, HRP linked AB (1:5000, 7074S, Cell Signaling), Anti-Mouse IgG HRP conjugate (1:5000, 96714S, Cell Signaling). Blots were developed by incubation with SuperSignal West Pico Plus Chemiluminescence Substrate (Thermo Scientific 34277) or/and SuperSignal West Atto Ultimate Sensitivity Substrate (Thermo Scientific A38555) and imaged using iBright CL1500 instrument (Thermo Scientific A44240)

*In-chip immunoassay:* Microfluidic channels were first primed with sterile filtered (0.2 um) PBS. Then the engineered nanocarriers were injected into the channel for 1 min, followed by serial rinsing with PBS and MiliQ for 3 min each. When necessary, a custom AH peptide (synthesized by Proteogenix) with the sequence SGSWLRDVWDWICTVLTDFKTWLQSKLDYKD was flown over the surface for 1 min at 200 µM in MiliQ water to rupture any remaining liposomes. The channel was then rinsed again for 3 min with PBS. All the functionalization steps were done at a flow rate of 50 µl/min by pipetting reagents sequentially to the inlet reservoir. The sample to be analyzed was then flown for 10 min at 5 µl/min and the channel was washed afterwards with PBS for 2 min to stop the assay and remove unbound analytes. The surface for the immunoassay was scanned over an area of 1000 × 100 $\mu m^2$ before and after the analyte was flown through the channel.

*Optical Setup:* For imaging we used a custom-built partially coherent digital in-line holography microscopy in a reflection-geometry with sample stabilization (SI Figure 11). The sample focus was actively stabilized via a proportional–integral–derivative loop based on the intensity distribution of the total internal reflection pattern from a 670 nm confocally illuminating beam (CPS670F, Thorlabs). For the illumination, light from 200-µm multimode fibre-coupled (M25L02, Thorlabs) 455 nm light emitting diode (M455F3 LED, Thorlabs) was outcoupled with a 6.24 mm aspheric lens (A110TM-A, Thorlabs) and 1:1 relay imaged onto the sample plane generated by a 1.42 NA oil immersion objective (UPLXAPO 60x, Olympus). In this configuration, a flat-top beam was delivered to the sample at an illumination numerical aperture of approximately 0.5. The partially coherent illumination limits the coherence length restricting detection to the vicinity of the glass–sample interface (< 5 µm). As the microfluidic channel height is approximately 100 µm, the opposing PDMS surface is well outside this range and therefore does not contribute to particle counts acquired at the glass sensing surface.[48] The beam covered a circular area 89.5 µm in diameter with a total power of 1.4–1.7 mW, corresponding to an irradiance of 0.22–0.27 $\mu W/\mu m^2$. For imaging, the light collected by the objective was directed through a 50:50 beamsplitter (BSW27, Thorlabs) and a 300 mm achromatic doublet (AC508-300A, Thorlabs) onto a scientific CMOS camera (C11440-22CU, 6.5 µm pixel size; Hamamatsu), providing an overall magnification of 100×. At the sample area, the microfluidic chip was mounted on a motorized XY microstage (Mad City Labs) with integrated linear encoders and an XYZ nanostage for high-precision positioning (Nano-LP200, Mad City Labs). Image acquisition was performed over a 66.6 µm × 66.6 µm field of view (1024 × 1024 camera pixels) with a 10 ms exposure time and a fixed frame rate of 100 Hz. The data were saved as 40 time-averaged frames resulting in an effective frame rate of 2.5 frames per second.

*Image Processing and Data Analysis:* We followed the same image processing and particle localization routines previously described.[31,49] All images were normalized framewise to the average background counts and then flatfield corrected to remove inhomogeneities along the optical system and imperfections in the sample illumination. Flatfield correction was applied by dividing the median value of the whole image scan dataset of each pixel to the corresponding pixel frame by frame. For particle localization, the global noise (σ) of each image was first estimated using the median absolute deviation, and candidate region of interests were then identified using a conservative detection threshold of 5σ. For detection of non-specific binding and quantify passivation against culture media and FBS (Figure 2), the detection threshold was lowered to 3σ. Binary images generated from the respective thresholds were then processed with morphological erosion to reduce noise artifacts. Particle candidate regions were further considered for localization if they satisfied both a connectivity (>2 pixels) and SNR threshold criteria within a 3×3 pixel neighborhood around the local extrema. Identified point-spread-functions were then segmented and sub-pixel localized using the radial symmetry centers algorithm,[50] yielding processed data sets as shown in SI Figure 12. Particle counts per condition were determined as average of counts in a $10\times10\ \mu m^2$ area and determined per each frame over the 1000 µm x 100 µm scan, ensuring both robust statistics and representative sampling. Frames containing clear artefacts such as damaged coverslip areas, as well as the first two and last two frames of a scan due to bleaching by prolonged LED exposure were discarded.

To relate the number of secreted EVs to cell confluency (see Figure 5d), cell images were taken with an inverted light microscope (Olympus CKX53, 4x) and cells within 3.15 mm × 2.36 mm regions of the well plate were counted using instance segmentation in Cellpose3.[51] For this, the pretrained Cyto3 model was fine-tuned on images acquired at the six inspected timepoints by iteratively correcting the segmentation masks. To evaluate the accuracy of the automated cell counts, eight smaller regions of interest (0.3 mm × 0.3 mm) were manually annotated. Each ROI contained between 31 and 91 cells. Manual counts were then compared with those obtained from the fine-tuned model. Differences between human and model counts ranged from 0 to 6 cells per frame, corresponding to a symmetric mean absolute percentage error of 5 ± 4 %. Most discrepancies occurred at ROI boundaries, where partial cell profiles were missed in the manual annotations but retained in the model's segmentation mask, which was computed on the full frame.

## Acknowledgements

P.R., S.S., C.L., J.O.A. and R.Q. acknowledge funding through the Swiss National Science Foundation grant 207485. The authors thank Miroslav Peterek and Justine Kusch of the Scientific Center for Optical and Electron Microscopy (ScopeM) at ETH Zurich for their support with CryoTEM imaging and FRAP measurements and Dr. Johannes Bader and Finn Brigger of the Drug Formulation & Delivery lab (ETH Zürich) for access to the NTA instrument.

## Conflict of Interest

The authors declare no competing interests.

## Author Contributions

Conceptualization: P.R., J.O.A., R.Q.; methodology: P.R., S.S., S.H.; investigation: P.R., S.S., S.H.; software: P.R, C.L., J.O.A.  formal analysis: P.R., C.L. visualization: P.R .; supervision: J.O.A., R.Q.; writing─original draft: P.R.; writing─review and editing: P.R., S.S., S.H., C.L., J.O.A., R.Q.